\documentclass[preprint,aps,prd,amsmath,amssymb]{revtex4}
\usepackage{graphicx}

\begin{document}
\title{Comments on the Chiral Symmetry Breaking in Soft Wall Holographic QCD}
\author{Jacopo Bechi}
\email[E-mail: ]{bechi@fi.infn.it}
\affiliation{CP$^{ \bf 3}$-Origins,
Campusvej 55, DK-5230 Odense M, Denmark.}
\begin{flushright}
{\it CP$^3$- Origins: 2009-28}
\end{flushright}
\begin{abstract}

In this paper we describe qualitatively some aspects of the holographic QCD. Inspired by a successfull 4D description, we try to separate the Confinement and the Chiral Symmetry Breaking dynamics. We also discuss the realization of the baryons like skyrmions in Soft Wall Holographic QCD and the issue of the Vector Meson Dominance.

\end{abstract}

\maketitle

\section{Introduction: the status of the art in bottom-up models}

The AdS/CFT correspondence provides a powerful tool to understand strongly coupled gauge theories. Both qualitatively and quantitatively also quite simple extra dimensional models of QCD motivated by the AdS/CFT correspondence \cite{Erlich:2005qh}\cite{Da Rold:2005zs}
have proved successful at reproducing some low energy hadronic data like vector and axial vector meson masses, decay constants and coefficients of the chiral Lagrangian. The AdS/QCD or, more generically, Holographic QCD models \cite{Gursoy:2007er}
 join together many older ideas which are useful for understanding features of the QCD at low energy as: hidden local symmetry, chiral symmetry breaking, large $N$ and Shifman-Veinshtein-Zakharov (SVZ) sum rules.

In general, in the bottom-up Holographic QCD models the starting point is to specify a 5D space time geometry and the fields that propagate there based on the proprieties of the QCD that we would like consider and on the usual AdS/CFT dictionary.

In the earlier AdS/QCD models the space is assumed to be $AdS_{5}$:

\begin{equation}
ds^{2}=g_{MN}dx^{M}dx^{N}=\frac{R^{2}}{z^{2}}(\eta_{\mu\nu}dx^{\mu}dx^{\nu}-dz^{2})
\end{equation}

where $\mu,\nu=(0,1,2,3)$ and $\eta_{\mu\nu}$ is the Minkowskian metric. Following the usual holographic rules, the $z$ coordinate ($z>0$) is dual at the energy scale $E$ from the 4D point of view $z=E^{-1}$. To mimic the confinement, an IR cutoff is taken at $z=z_{m}$ and the towers of the Kaluza-Klein (KK) modes are identified with the towers of the radial excitations of the QCD mesons.

Despite their success, these simplest models have some shortcomings; for examples the behavior of the masses of the meson resonances is $m^{2}_{n}\sim n^{2}$ instead of the phenomenological Regge behavior $m^{2}_{n}\sim n$. In an attempt to obtain more realistic features of the excited mesons the AdS/QCD model can be modified to include a dilaton with a quadratic profile in IR \cite{Karch:2006pv}
. This class of models are called Soft Wall AdS/QCD models.
However, while the correct Regge spectrum of the resonances is indeed produced with this Soft Wall, problems raise about the chiral symmetry breaking. More precisely, the chiral symmetry breaking, signaled by a not trivial background profile of the 5D scalar field dual to the quark bilinear $\bar{q}q$, is not QCD like because the explicit and the spontaneous breaking are not independent. This problem has been recently considered by \cite{Gherghetta:2009ac} that, adding a quartic term to the scalar potential decouples the quark mass from the chiral condensate. The 5D action of \cite{Gherghetta:2009ac} is

\begin{equation}\label{gherghetta}
S_{5}=-\int d^{5}x\sqrt{-g}e^{-\phi(z)}\text{Tr}\Big[|DX|^{2}+m_{X}^{2}|X|^{2}-\kappa|X|^{4}+\frac{1}{g_{5}^{2}}(F_{L}^{2}
+F_{R}^{2})\Big]
\end{equation}

and the background is the full $AdS_{5}$ with $z \in (0,\infty)$. In \eqref{gherghetta} the field tensor $F_{L,R}$ are the field tensor of the 5D gauge group $SU(2)_{L}\times SU(2)_{R}$ with gauge potentials respectively $A_{L}$ and $A_{R}$, and

\begin{equation}
D^{M}X=\partial^{M} X-iA^{M}_{L}+iXA_{R}^{M}
\end{equation}

where $X(x,z)$ is dual to the operator $\bar{q}q(x)$ and $m_{X}^{2}=-3/R^{2}$. The paper \cite{Gherghetta:2009ac} considers also the issue of the chiral symmetry restoration for highly excited mesons. The earlier Soft Wall models predict that the chiral symmetry is restored for highly excited mesons but, other studies as \cite{Shifman:2007xn}, suggest the opposite. Indeed the highly excited resonance should be sensible to the confining dynamics and, in a large $N$ analysis, has been proved, with quite general assumptions, by Coleman-Witten Theorem \cite{Coleman:1980mx}
 that the confinement necessary induces the breaking of the chiral symmetry to the diagonal subgroup. An holographic incarnation of this theorem can be seen for example in the famous Sakai-Sugimoto model \cite{Sakai:2004cn}
 in which the fusion of the $D8$ with the $\overline{D8}$ flavor branes is inevitable in the confining background produced by the $D4$ color branes. In \cite{Gherghetta:2009ac} this problem is solved taking a background profile $\langle X(z)\rangle=v(z)$ linear in IR, that is

\begin{equation}
v(z\rightarrow \infty)\sim z
\end{equation}

and solving the scalar equation of motion for the dilaton field $\phi(z)$ that gives an IR quadratic behavior for the dilaton.

The results obtained by \cite{Gherghetta:2009ac} about the meson spectrum are very good with the only exception of the $\rho$ meson that gets a too low mass. The baryons are not discussed.

In despite of the remarkable phenomenological success of the model proposed in \cite{Gherghetta:2009ac} some aspects are put by hand, as for example the quartic term in the bulk scalar potential. The aim of this paper is to study the issue of the chiral symmetry breaking in AdS/QCD Soft Model considering as closely as possible some dynamical aspects about the QCD at low energy that has been proved successful in 4D. Our paper is based on some ideas before proposed in \cite{Shuryak:2007uq}
 that here we try to develop.

\section{A picture of the QCD vacuum}

 In QCD there are strong indications that the Chiral Symmetry Breaking and the Confinement are due to two different dynamical mechanisms. The confinement is quite mysterious still, but a widely accepted paradigm is that it is a dual Meissner effect related to the formation of cromoelectric string in a magnetic monopole condensate. For the Chiral Symmetry Breaking a very successfull model exists, the Instanton Liquid Model (ILM) (for a review \cite{Schafer:1996wv}). In this model the most important dynamical object is the instanton and the quarks get mass because the zero modes hop between instantons changing chirality. In other terms, the hopping of the quark zero modes between instantons provide a mechanism to break the Chiral Symmetry Breaking by meaning of the formation of a collective vacuum state. In this model the classical meson physics is due to the quark zero mode dynamics.

\section{Quark zero modes in AdS/QCD and the vacuum state}

In the precedent section we have said that the instanton dynamics and the quark zero modes are the crucial ingredients to understand the Chiral Symmetry Breaking in QCD. So is interesting to try to rewrite this phenomena in an holographic language with the aim to build holographic models more possible closed to the real QCD. In \cite{Bechi:2009dy}
 we have studied the problem of the quark zero mode creation in AdS/QCD. In AdS space the 4D instanton is dual to a event like object in the bulk, the D-instanton, localized at $z=\rho$, where $\rho$ is the size of the instanton. This D-instanton is electrically coupled to a 0-form $C_{0}$ dual to the 4D topological charge density $\text{Tr}F_{\mu\nu}F_{\alpha\beta}\epsilon^{\mu\nu\alpha\beta}$, where the trace is over color indices.

The gauge group in the bulk is $U_{L}(N_{F})\times U_{R}(N_{F})$. We assume that also the axial symmetry $U_{A}(1)$ is gauged because this is a symmetry of the perturbative QCD in absence of topologically non trivial configurations. To generate the quark zero modes in presence of a instanton is necessary to add to the 5D action the following term permitted by the symmetries

\begin{eqnarray}\label{new}
&&S_{new}=\alpha\int C_{3}\wedge \text{Tr}(F_{L}-F_{R})=\sum^{N_{F}}_{a=1}\int C_{3}\wedge (F_{L}^{aa}-F_{R}^{aa}),\nonumber\\
&&F=(F)^{ab}\qquad a,b=1,\dots,N_{F}.
\end{eqnarray}

In \eqref{new} $C_{3}$ is defined by means of the Hodge duality $dC_{0}=H_{1}=\star H_{4}=\star dC_{3}$. The minus sign between the left and the right tensor is demanded by the discrete symmetries of QCD. The parameter $\alpha$ can be fixed by matching with Axial Anomaly.

With this term added, is possible to show the quark zero mode creation effect in the bulk of AdS. Generally in AdS we can work only with color invariant quantity and so we have not the usual description of the quark zero modes in term of the quark fields $\psi_{0}(x)$ like in 4D because these fields are not gauge invariant. Anyway, a gauge invariant signal of the creation of a zero modes is given by the production of the currents $j_{L,R\mu}^{aa}=\bar{\psi}_{L,R}\gamma_{\mu}\tau_{L,R}^{aa}\psi_{L,R}$, where $\tau_{L,R}^{aa}$ indicates the generator of the corresponding $U_{L,R}(1)$ Abelian subgroup of $U_{L}(N_{F})\times U_{R}(N_{R})$. Indeed, these currents, that are dual to the $A_{L,R\mu}^{aa}$ in AdS are excited by the $H_{1}$ field product by the D-instanton. The fields product by a D-instanton can be absorbed by a anti D-instanton and this is dual to the delocalization of the quark zero modes in 4D. In 4D this effect leads to the formation of the collective state, signaled by the formation of the chiral condensate $\langle\bar{q}q\rangle$, that breaks the chiral symmetry $SU_{L}(N_{F})\times SU_{R}(N_{F})$ to the diagonal Isospin group. In 5D the production of the quarks zero modes motivate us to introduce in the bulk a $D3$ brane like object that represents the non-perturbative vacuum; we will call this dynamically formed object $D3$ brane even if it is not a brane in the sense of a BPS state in string theory because the theory is not supersymmetric. At the best of my knowledge, the first paper (and also the only) to propose to represent the chiral symmetry breaking in AdS/QCD by using a dynamically formed extended object has been \cite{Shuryak:2007uq}
. The $D3$ brane has a thickness in the $z$ direction between $z_{min}\simeq (1\quad GeV)^{-1}$ and $z_{max}\simeq (200/300\quad MeV)^{-1}$. To simplicity we consider the $D3$ brane without thickness.

The dynamics of the $D3$ brane is described by the Nambu-Goto action

\begin{equation}\label{azionebrana}
S_{D3}=-f^{4}\int_{V_{4}}d^{4}x\sqrt{|g_{ind}|}
\end{equation}

where $x$ are the coordinate on the worldvolume $V_{4}$, $\sqrt{|g_{ind}|}$ is the absolute value of the determinant of the induced metric of $D3$ and $f$ indicated the its tension. In \eqref{azionebrana} we are neglecting a dilaton depending potential that provides to the equilibrium position $\bar{z}$ of the brane. We leave this problem for future work and treat the equilibrium position $\bar{z}$ as a free parameter of the model.

The $D3$ brane shape is described by the embedding $X^{M}(x)$ and, by using the parametrization invariance, we can get $X^{M}=(x^{\mu},Z(x))$. Remembering that the induced metric is

\begin{equation}
(g_{ind})_{\mu\nu}=\partial_{\mu}X^{M}\partial_{\nu}X^{N}g_{MN}(X)
\end{equation}

we get

\begin{equation}
(g_{ind})_{\mu\nu}=\frac{R^{2}}{Z^{2}(x)}(\eta_{\mu\nu}-\partial_{\mu}Z(x)\partial_{\nu}Z(x))
\end{equation}

Now let's take $Z(x)=\bar{z}+\tilde{Z}(x)$ and work at the first order in the approximation $\partial_{\mu}Z\ll 1$ and $\tilde{Z}(x)\ll \bar{z}$, from which

\begin{equation}
\sqrt{|g_{ind}|}\simeq \frac{R^{4}}{\bar{z}^{4}}(1-\frac{1}{2}\partial_{\mu}Z\partial^{\mu}Z)
\end{equation}

and, putting in \eqref{azionebrana} and neglecting a constant term, we get

\begin{equation}\label{gigio}
S_{D3}=f^{4}\frac{R^{4}}{\bar{z}^{4}}\int d^{4}x \frac{1}{2}\partial_{\mu}\tilde{Z}\partial^{\mu}\tilde{Z}.
\end{equation}


The field $\tilde{Z}(x)$ is related to the scalar glueball with the quantum number of the scale anomaly which, in absence of a mechanism that stabilize the brane position is the Goldstone boson of the spontaneous breaking of the scale invariance. The treatment of the scalar sector is, in general, problematic and it will not be discussed in this paper. We set the position of the brane at the its equilibrium value $\bar{z}$ and consider this as a free parameter of the model. Should be remark that the aim of this paper is not to improve quantitatively the holographic QCD model. Because QCD is deprived of large hierarchy and the only physically relevant scale is $\Lambda_{QCD}$ is clear that also models that does not describe precisely the real QCD dynamics can give good phenomenological prediction. However, to recognize the existence of different dynamics at the base of the chiral physics and of the confinement physics (although not very separate in energy), is conceptually important and can bring simplifications. So the our goal is just to try to build an holographic description of QCD that be closer as possible at the known dynamics of QCD.

Considering, for simplicity, the case $N_{F}=2$, we introduce a complex field $\Phi(x)$, localized at $z=\bar{z}$, with the quantum number of the chiral condensate $\langle \bar{q}_{L}q_{R}\rangle$. The dynamical motivation to introduce this degree of freedom is the zero mode production by D-instantons. This fields represent at low energy the collective state due to the delocalization of the zero modes between the topological charges and which breaks the chiral symmetry (see \cite{Fabbrichesi:2008ga} for an Holographic Electroweak Symmetry Breaking model in which a localized scalar field is also introduce in the bulk). We use the usual parametrization of the degree of freedom at low energy

\begin{equation}
\Phi(x)=\Big(v+\sigma(x)\Big)U(x)
\end{equation}

where $U(x)$ is a $SU(2)$ matrix that, respect to the bulk chiral group $SU(2)_{L}\times SU(2)_{R}$, transform like $U(x)\rightarrow g_{L}(x, \bar{z})U(x)g_{R}^{\dag}(x, \bar{z})$ and to represent the coset space of the chiral symmetry breaking (that is to say that $\langle U(x)\rangle =\mathbf{1}$) and the V.E.V. $v$ is a free parameter related at the mass difference between the vector $\rho$ and the axial vector $a_{1}$ meson. The action for $\Phi(x)$ is the usual

\begin{equation}\label{effettiva}
S_{\Phi}=\int d^{4}x \frac{1}{2}\text{Tr}\big[(D_{\mu}\Phi)^{\dag}D^{\mu}\Phi+V(\Phi^{\dag}\Phi)\Big]
\end{equation}

with $D_{\mu}=\partial_{\mu}+iA_{\mu}^{L}-iA_{\mu}^{R}$. The $\sigma$ meson physics is sensible about the detail of the potential $V(\Phi^{\dag}\Phi)$. Phenomenologically we could consider the potential parameters to be fixed by the comparison with the experimental data. However in this paper we neglect the scalar meson physics and set $\Phi(x)=vU(x)$.

To conclude this section, it is worth to spent some words to remark the most important difference respect to the usual AdS/QCD models. The first difference is the introduction, dynamically motivated, of a scalar field localized in the bulk that represent the chiral condensate. This is particularly important in the soft wall model where, as discussed in the first section, the implementation of the chiral symmetry breaking is not direct. Another, conceptual, difference respect at the usual is the understanding of the brane at $z=\bar{z}$. In the approach described in this paper this brane is related at the quark zero modes. Several indication from ILM strongly suggest that the physics of the classical meson is saturate considering only the contribution of the zero modes. So we have to take the position of this brane as setting the typical size of the classical meson. If we use a Kaluza-Klein (KK) approach taking $\bar{z}$ as size of the extra dimension we have to retain only the first KK resonance and consider them to be the classical meson. Contrary, in the usual Hard Wall approach the IR brane is introduced not like a dynamical object and is seen as an incarnation of the confinement. From this point of view the KK tower is regarded as the tower of the radial excitation of the meson. So the our approach is, in some sense, between the Hard Wall models and the Soft Wall models.

\section{Skyrmion in Soft Wall AdS/QCD}

As already discussed, the approach proposed in this paper, although theoretically interesting, unlikely can meanly improved the numerical results obtained with other models in the meson sector.

However the our approach can be maybe more useful in the baryon sector. On one hand, the ILM gives us evidence that also the physics of the light baryons is understandable in term of the quark zero modes only. That is as to say that the confining potential is not important for the quark binding inside the nucleon. On the other hand, it is well known that, in a effective approach to the low energy QCD in term of meson field, the baryon can be successfully described as a soliton made of mesons fields. This soliton is usually referred as skyrmion.

In the last years the holographic realization of the skyrmion has been studied both in the up-down stringy approach \cite{Nawa:2006gv}\cite{Hong:2006ta}\cite{Hata:2008xc}
\cite{Hashimoto:2008zw}
and in bottom-up hard wall approach \cite{Pomarol:2007kr}\cite{Pomarol:2008aa}
. The two ways are different respect to the mechanism that stabilizes the skyrmion. In the up-down models the Derrick's instability is avoid because, automatically, the model has a Skyrme term. However, we are more interested to the case of bottom-up models that now we review briefly.

At the best of my knowledge, the first papers to treat extensively this topis were \cite{Pomarol:2007kr}\cite{Pomarol:2008aa} (but see also \cite{Son:2003et}
 where the problem is considered in a pioneer deconstructed version of holographic QCD).

In \cite{Pomarol:2008aa} the model consists of a 5D $U_{L}(2)\times U_{R}(2)$ gauge theory on a slice of AdS space with boundaries $[z_{UV},z_{IR}]$. The chiral symmetries is broken to the diagonal subgroup by the boundary conditions on the boundary at $z=z_{IR}$. The $U(2)^{3}$ anomaly of QCD is reproduced holographically adding a Chern-Simons term

\begin{equation}
S_{CS}=-i\frac{N_{c}}{24\pi^{2}}\int [\omega_{5}^{L}-\omega_{5}^{R}].
\end{equation}

Then, the skyrmion is identified with time independent topologically non trivial configuration labeled by the topological charge

\begin{equation}
B=\frac{1}{32\pi^{2}}\int d^{3}x\int^{z_{IR}}_{z_{UV}}dz \epsilon_{\hat{\mu}\hat{\nu}\hat{\rho}\hat{\sigma}}\text{Tr}\Big[L^{\hat{\mu}\hat{\nu}}L^{\hat{\rho}\hat{\sigma}}-
R^{\hat{\mu}\hat{\nu}}R^{\hat{\rho}\hat{\sigma}}\Big]
\end{equation}

where the indices $\hat{\mu},\hat{\nu},\dots$ run over the 4 spatial coordinates. In this approach the Chern-Simons term is crucial to stabilize the skyrmion size. The symmetry breaking boundaries condition are crucial for $B$ to be quantized and to cancel the gauge variation of the Chern-Simons term on the IR boundary. From a 4D point of view this skyrmion is composed by the pion field (that can be introduced like a axion \cite{Panico:2007qd}) and the entire tower of vector and axial vector meson resonances.

This model is very interesting and the analysis of the several baryon proprieties is in good accord with the observations. Nevertheless it is based on the hard wall approach and so seems worth to try to get a similar description in a soft wall model. Now we try to do it using the model introduced before.

The model that we consider in this section is builded with the following ingredients:

\begin{itemize}

\item $SU_{L}(2)\times SU_{R}(2)\times U_{B}(1)$. We don't introduce the axial symmetry $U_{A}(1)$ because this symmetry is anomalous. However, in the large $N$ limit the mass of the $\eta'$ is proportional to $1/N$ and one could want include this meson in the low energy description. We don't do so here.

\item The boundary conditions (for the lighter KK resonances) at $z=\bar{z}$ are Neumann for all the gauge fields. That is to say that the boundary conditions don't break the chiral symmetry; this symmetry is broken only by the V.E.V. of the scalar field at $z=\bar{z}$.

\end{itemize}

Besides the term \eqref{effettiva} there are other terms that we have to consider for the scalar action on the brane.

The first term is

\begin{equation}
S_{Bar.}=\int d^{4}x B_{\mu}(x,\bar{z})J^{\mu}_{top.}(x)
\end{equation}

where $J^{\mu}_{top.}(x)$ is the topological baryon current \cite{Witten:1983tw}

\begin{equation}
J^{\mu}_{top.}=\frac{1}{24\pi^{2}}\epsilon^{\mu\nu\alpha\beta}\text{Tr}\Big[U^{-1}\partial_{\nu}UU^{-1}\partial_{\alpha}U
U^{-1}\partial_{\beta}U\Big].
\end{equation}

This term is very important for the stability of the skyrmion of the field $U(x)$. Indeed is known that, in the Hidden Local Symmetry approach at which the model that we are studying is very closed, the stabilization due to the vector meson $\rho$ is canceled by the axial vector meson $a_{1}$. Then the stability of the skyrmion is guaranteed by the $\omega$ meson, introduced gauging the baryon number symmetry $U_{B}(1)$, like studied in \cite{Adkins:1983nw}
.

In the case of only two flavor, the Wess-Zumino-Witten (WZW) \cite{Wess:1971yu}\cite{Witten:1983tw} term doesn't exist. Mathematically this is because the Homotopy Group $\pi_{4}(SU(2))$ is not trivial. From a more physical point of view it is due to absence of $SU(2)^{3}$ anomalies. However, also in the case $N_{F}=2$, we have to consider the anomalous decay $\pi^{0}\rightarrow \gamma\gamma$. So we promote $U(x)$ to a $SU(3)$ matrix, insert in the action the WZW term and then reduce to only the $SU(2)$ degree of freedom. We introduce the gauged WZW term on the $D3$ brane.

At this point the Lagrangian on the $D3$ brane, that represent QCD effective theory below $E\sim 1/\bar{z}$, is a 4D theory with a non-linearly realized $SU_{L}(2)\times SU_{R}(2)$ in which all the group is weakly gauged with gauge coupling $g_{eff}\propto g_{5}$. The point on which we want to put emphasis here is that trying to stay as close as possible at the known dynamics of QCD the holographic description simplify considerably and that emerges is a geometrical embedding of several old ideas.

To conclude we want spend some words about possible inconsistencies of the theory due to bulk gauge anomalies. Both $S_{top}$ and the WZW term are anomalous under a bulk gauge variation. Indeed these terms are added to the effective Lagrangian to fulfil the Wess-Zumino matching condition and to reproduce at the composite level the anomaly at the quark level. Now the anomalous symmetry are gauged so one could wonder if the 5D theory is in trouble. Of course if we want to consider the 5D theory as a complete quantum theory it is not consistent. But we have to remember that, differently from AdS/CFT, the AdS/QCD models have to be understood as effective models valid at the classical level. Indeed, from a 4D point of view, the gauge fields on the $D3$ brane (that approximatively coincide with the KK resonance) are dual to the spin one meson. Like in the Hidden Local Symmetry approach this fields are composite massive fields and both the compositeness scale and the mass scale are $\sim 1/\bar{z}$. The anomaly in the bulk is localized at $\bar{z}$ so at the scale of the compositeness of the meson. But the composite meson are good degree of freedom only below this scale. The anomaly is proportional to $F\wedge F$ but, because the gauge boson are massive, the contribution from the composite gauge degree of freedom with frequencies below the mass is suppressed. So this anomaly is tolerable from a 4D effective point of view. In this way we implement the Vector Meson Dominance (VMD).

If we want to preserve the gauge symmetry in the bulk we have to give up the complete VMD and to allow to a direct coupling between photon and pion. Indeed we can use the non-local degree of freedom $\tilde{U}(x)$

\begin{equation}
U(x)\rightarrow \tilde{U}(x)=P\Big[\exp\Big(-i\int_{z_{UV}}^{\bar{z}}dz' R_{5}(x,z')\Big)\Big]U(x)P\exp\Big[\Big(i\int_{z_{UV}}^{\bar{z}}dz' L_{5}(x,z')\Big)\Big]
\end{equation}

to describe the pion degree of freedom. The symmetry transformation of $\tilde{U}(x)$ is

\begin{equation}
\tilde{U}(x)\rightarrow g^{\dag}_{L}\tilde{U}(x)g_{R}
\end{equation}

where $g_{L}$ and $g_{R}$ are transformation at $z=0$. The Electroweak Gauge Symmetry can be introduced gauging a subgroup of these transformations. So, if we want to preserve the bulk gauge symmetry, we have to introduce a non locality in the $z$ direction that, by a 4D point of view correspond to a non locality in the energy. This non locality in the energy is expected because $\tilde{U}(x)$ is a degree of freedom of the effective theory at low energy.

The non locality in the extra dimension is not less ugly of the gauge anomaly but become acceptable if the 5D model is understood just like a low energy effective model of 4D physics.

The fields $\tilde{U}(x)$ and $U(x)$ are gauge equivalent from a 5D point of view so the holographic model can not predict if one have to implement the complete VMD or not.

\section{Conclusion}

In this paper we have addressed qualitatively some aspects of the holographic description of QCD trying to stay as close as possible to some successfull 4D dynamical descriptions. The collective state formed by quark zero modes is introduced in 5D like a $D3$ brane like extended object. While we don't expect relevant changes in the quantitative prediction of the model, some aspect, as the implementation of phenomena related to the Chiral symmetry Breaking in Soft Wall models, could become more clear.

\vspace{1cm}

\textbf{Acknowlegdements}: This paper has been written when I was guest at the $CP^{3}-Origins$ that I thank for hospitality. Also I would like to thank the Fondazione Angelo della Riccia for financial support. I thank F. Sannino and especially J. Schechter for discussions.


\end{document}